\documentclass[doublecol]{epl2}
\usepackage{graphicx}

\newcommand{\kb}{k_\mathrm{B}}
\newcommand{\e}{{\rm e}}
\renewcommand{\d}{{\rm d}}

\newcommand{\xmaa}{x_\mathrm{A}}
\newcommand{\xmab}{x_\mathrm{B}}
\newcommand{\xmac}{x_\mathrm{C}}
\newcommand{\xmad}{x_\mathrm{D}}
\newcommand{\xmi}{x_\mathrm{m}}

\newcommand{\ymaa}{y_\mathrm{A}}
\newcommand{\ymab}{y_\mathrm{B}}
\newcommand{\ymac}{y_\mathrm{C}}
\newcommand{\ymad}{y_\mathrm{D}}
\newcommand{\ymi}{y_\mathrm{m}}
\newcommand{\pder}[2]{\frac{\partial #1}{\partial #2}}
\renewcommand{\vec}[1]{\mbox{\boldmath $#1$}}

\newcommand{\mf}{{\rm m}_1}
\newcommand{\mt}{{\rm m}_2}

\title{A Reversibility Parameter for a Markovian Stepper}
\author{T. Harada \inst{1} \and N. Nakagawa \inst{2}}
\institute{
	\inst{1} Department of Human and Artificial Intelligent Systems, University of Fukui, Fukui 910-8507, Japan \\
	\inst{2} Department of Mathematical Sciences, Ibaraki University, Mito 310-8512, Japan
}
	

\abstract{
Recent experimental studies on the stepwize motion of biological molecular motors have revealed that the ``characteristic distance'' of a step is usually less than the actual step size.
This observation implies that the detailed-balance condition for kinetic rates of steps is violated in these motors.
In this letter, in order to clarify the significance of the characteristic distance, we study a Langevin model of a molecular motor with a hidden degree of freedom.
We find that the ratio of the characteristic distance to the step size is equal to unity if the dominant paths in the state space are one dimensional, while it deviates from unity if the dominant paths are branched.
Therefore, this parameter can be utilized to determine the reversibility of a motor even under a restricted observation.
}

\begin{document}

\maketitle

Single-molecule measurement techniques have expanded the possibilities for studying biological macromolecules from physical points of view.
For instance, molecular motors, which move along protein filaments in a stepwise manner fueled by the hydrolysis of adenosine tri-phosphate (ATP), have been investigated extensively.
Through many studies, several of the important properties of molecular motors, including their step sizes and the kinetic rates of their forward and backward steps, have been experimentally determined.

In this letter, we study the modulation of the kinetic rates by an external force.
Such phenomena have been examined in recent experiments \cite{Nishiyama:2002, Carter:2005, Taniguchi:2005, Hirakawa:2000, Clemen:2005}.
Interestingly, the results of these experiments suggest that the detailed-balance (DB) condition for the kinetic rates is violated in the stepwise motion of a motor.

Here, we start from the DB condition for a system that has several discrete states \cite{Hill:1977}.
Let $\mf$ and $\mt$ denote two  of these states, and let $k_{21}$ and $k_{12}$ represent the transition rates from $\mf$ to $\mt$ and vice versa, respectively.
Then, the DB condition is given as
\begin{equation}
\frac{k_{21}}{k_{12}} = \exp(\beta \Delta G_{12}),
\label{e.db}
\end{equation}
where $\beta \equiv (\kb T)^{-1}$ is the inverse temperature of the heat bath ($\kb$ being the Boltzmann constant), and $\Delta G_{12} \equiv G_1 - G_2$ represents the difference in the (free) energy between the states $\mf$ and $\mt$.

Note that the DB condition given in eq.~(\ref{e.db}) is derived from a more microscopic condition termed the local detailed-balance (LDB) condition with several assumptions.
Let $\vec{r}_1$ and $\vec{r}_2$ denote microstates in the states $\mf$ and $\mt$, respectively, and let $[\vec{r}]$ represent a microscopic path connecting $\vec{r}_1$ and $\vec{r}_2$ in an interval $t$.
Let $P([\vec{r}] | \vec{r}_1)$ represent the transition probability of a path $[\vec{r}]$ with given $\vec{r}_1$.
Then, the LDB condition is expressed as
\begin{equation}
\frac{P([\vec{r}] | \vec{r}_1)}{P([\tilde{ \vec{r}} ] | \vec{r}_2)} = \exp \left[\beta \Delta G(\vec{r}_1, \vec{r}_2) \right],
\label{e.ldb}
\end{equation}
where $[\tilde{\vec{r}}]$ is the time-reversed path of $[\vec{r}]$, and $\Delta G(\vec{r}_1, \vec{r}_2) \equiv G(\vec{r}_1) - G(\vec{r}_2)$ represents the difference in the (free) energy between the microstates $\vec{r}_1$ and $\vec{r}_2$.
Eq.~(\ref{e.db}) is derived from eq.~(\ref{e.ldb}) provided that the thermal energy is sufficiently less than the hight of a barrier separating $\mf$ and $\mt$, and the curvature of $G(\vec{r})$ around these states are the same.
It has been known that the LDB condition is an essential condition in order to make a stochastic model compatible with thermodynamics, and to derive several nonequilibrium equalities, including the fluctuation theorems and the Jarzynski equality \cite{Maes:1999, Crooks:2000}.
Furthermore, it has been argued that eq.~(\ref{e.ldb}) is satisfied in a wide class of models, including Markov chains, Langevin equations, and deterministic systems connected to heat baths \cite{Maes:1999, Crooks:2000, Harada:2006, Jarzynski:2000}.

When one wants to test the validity of eq.~(\ref{e.db}) for a given system, one has to know $\Delta G_{12}$. 
From experimental point of view, this is sometimes difficult.
Instead, eq.~(\ref{e.db}) can be indirectly verified by changing the magnitude of the external force $\vec f$ applied to the system.
If the magnitude of the external force is changed by a small amount $\vec{\delta f}$, $\Delta G_{12}$ would be modified as $\Delta G_{12} \to \Delta G_{12} + \vec{\delta f}\cdot\vec{\ell}$, where $\vec{\ell}$ is a vector connecting $\mf$ and $\mt$.
Then, according to eq.~(\ref{e.db}), the ratio of the kinetic rates would be modified as
\begin{equation}
\frac{k_{21} (\vec f + \vec{\delta f})}{k_{12} (\vec f + \vec{\delta f})} = \frac{k_{21}(\vec f)}{k_{12} (\vec f)} \exp \left( \beta \vec{\delta f} \cdot \vec{\ell} \right).
\label{e.mdb}
\end{equation}
Note that this condition can be verified without determining $\Delta G_{12}$.

In the case of a molecular motor, the motor exhibits stepwize motion along a one-dimensional filament with a fixed step size $\ell$.
It would be possible to view this behavior as transitions among discrete states.
Then, one would expect the condition given in eq.~(\ref{e.mdb}) holds for the kinetic rates of the forward and backward steps of the motor.
Actually, it has been used in one-dimensional hopping models as a constraint for the transition rates in the model \cite{Fisher:1999, Kolomeisky:2005, Seifert:2005}.
However, single molecule-experiments revealed that eq.~(\ref{e.mdb}) does not hold for a variety of molecular motors\cite{Nishiyama:2002, Carter:2005, Taniguchi:2005, Hirakawa:2000, Clemen:2005}.

For instance, Nishiyama {\it et al.}~measured the kinetic rates of forward and backward steps of conventional kinesin, denoted by $k_+$ and $k_-$, respectively, for varying magnitude of the external force $f$ applied to the motor in parallel to the filament. They obtained data implying
\begin{equation}
\frac{k_+ (f + \delta f)}{k_- (f + \delta f)} = \frac{k_+(f)}{k_- (f)} \exp \left( \beta \delta f \ell \theta \right),
\label{e.mdbr}
\end{equation}
where $\theta$ is a dimensionless parameter, whose value was $\theta \simeq 0.4$.
This result has been confirmed by subsequent studies \cite{Carter:2005, Taniguchi:2005}.
Furthermore, the value of $\theta$ has been measured for several other species of molecular motors, including 22S dynein (taken from \textit{Tetrahymena} cilia) \cite{Hirakawa:2000} and myosin V \cite{Clemen:2005}, and the results in those cases also indicate that the value of $\theta$ is significantly less than unity.

Conventionally, the dimensionless quantity $\theta$ introduced above has been referred to as the ratio of the sum of the ``characteristic distances'', $d$, of forward and backward steps to the actual step size $\ell$: $d = \theta \ell$ \cite{Nishiyama:2002, Taniguchi:2005, Bell:1978}.
Thus, the above experimental results imply that the characteristic distance is less than the actual step size of molecular motors. However, the physical significance of the characteristic distance has not been clarified yet.

These experimental findings imply that eq.~(\ref{e.mdb}), and thus eq.~(\ref{e.db}), cannot be applied to the kinetic rates of the forward and backward steps of molecular motors.
This empirical observation, however, may not suggest that the LDB condition [eq.~(\ref{e.ldb})], from which eq.~(\ref{e.db}) is derived, is not applicable to microscopic models of a molecular motor.
Actually, through investigations of the Brownian motor models \cite{Julicher:1997, Reimann:2002}, it has been realized that the kinetic rates $k_\pm$ does not satisfy eq.~(\ref{e.db}).
However, the mechanism of violation of eq.~(\ref{e.db}) in these models has not been fully elucidated. This is because it is in general not easy to obtain simple expressions of the kinetic rates, $k_\pm$, for the conventional Brownian motor models.
Furthermore, in the earlier studies of the Brownian motor models, the LDB condition has not been always implemented.

In this letter, we present a simple model of a molecular motor that satisfies the LDB condition and for which $k_\pm$ can be explicitly calculated.
For this model, we demonstrate that the dimensionless parameter $\theta$ introduced above actually deviates from unity, and we clarify the mechanism of this deviation.
This also clarifies the physical significance of the characteristic distance $d$ of a molecular motor.
Finally, we demonstrate that the dimensionless parameter $\theta$ provides a good measure to indicate the structure of the paths in the state space of the system.

We here study a model possessing a ``hidden'' degree of freedom that is neither observed nor controlled experimentally.
For example, this hidden variable could represent the extent of ATP hydrolysis reaction, i.e., the degree of completion of the reaction, which is not monitored in most experiments.
This ``hidden'' degree of freedom is also conceptually equal to the internal state that is often incorporated into a Brownian motor model.
Here, for simplicity, we assume that there is only one such degree of freedom, denoted by $y$, and that its timescale is comparable to that of the spatial degree of freedom, $x$.

Employing the above-stated assumptions, we study the following set of Langevin equations:
\begin{eqnarray}
\gamma \dot x(t) &=& - \pder{U(x(t), y(t))}{x} + f + \xi(t), \label{e.2d1}\\
\Gamma \dot y(t) &=& - \pder{U(x(t), y(t))}{y} + F + \Xi(t). \label{e.2d2}
\end{eqnarray}
Here, $\gamma$ and $\Gamma$ are the friction coefficients for $x$ and $y$, $\xi(t)$ and $\Xi(t)$ are zero-mean white Gaussian noises whose variances are $2 \gamma \kb T$ and $2 \Gamma \kb T$, respectively, $f$ represents an external force applied to the particle, and $F$ represents a driving force conjugate to $y$.
We impose a certain translational symmetry on the potential $U(x, y)$, which is specified below, so that the motion of the particle is $\ell$-periodic when projected onto the $x$ axis. We also assume that the motion of the particle is $L$-periodic when projected onto the $y$ axis, noting that the ATP hydrolysis reaction occurs in a cyclic manner, and we impose an appropriate translational symmetry on the potential.
With this interpretation, the product of the driving force, $F$, and the periodicity of the system along the $y$ direction, $L$, is interpreted as the change in the chemical potential occurring in a single ATP hydrolysis reaction: $\Delta \mu=FL$. Then, the total free energy is defined as $V(x, y) = U(x, y) - f x - F y$.
This type of model has been investigated as a model of molecular motors in several contexts \cite{Magnasco:1994, Terada:2002}.
Furthermore, by interpreting $y$ as an internal state of the motor, this model can be connected to standard Brownian motor models with discrete internal states \cite{Julicher:1997, Reimann:2002} by considering appropriate form of the potential $U(x, y)$ and letting $\Gamma/\gamma \ll 1$.
An important property of the system described by eqs.~(\ref{e.2d1}) and (\ref{e.2d2}) is that it satisfies the LDB condition [eq.~(\ref{e.ldb})].
This can be easily confirmed by use of Onsager-Machlup path integral representation \cite{Onsager:1953, Harada:2006}.

\begin{figure*}
\onefigure{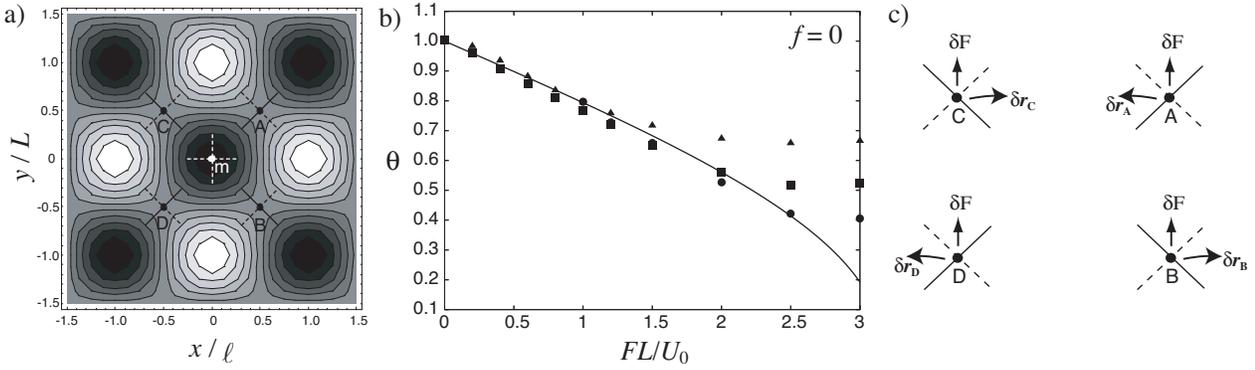}
\caption{a) Form of the potential $U(x, y) = - U_0 \cos(\pi x / \ell) \cos(\pi y / L)$ (Model I) represented by a grayscale, with the value of $U$ decreasing as the shade becomes darker. The small white dot labeled \textbf{m} denotes a minimum, and the small black dots labeled \textbf{A}, \textbf{B}, \textbf{C} and \textbf{D} denote saddles. The dashed lines (solid lines) associated with the fixed points indicate the stable (unstable) directions with positive (negative) eigenvalues of the Hessian matrix, $\mathcal{H}$. b) The reversibility parameter $\theta$ as a function of the driving force $F$ with $f = 0$. The solid curve represents the value of $\theta$ calculated using eq.~(\ref{e.simpletheta}). The solid triangles, squares and circles represent the values of $\theta$ calculated through numerical simulation of the Langevin equations (\ref{e.2d1}) and (\ref{e.2d2}) for $\beta U_0 = 5, 10$ and $20$, respectively. The error bars are smaller than the symbols. c) Schematic representation of the shift of the saddles under a perturbation, $\delta \vec{E} = (0, \delta F)$ (indicated by small arrows). The arrows $\delta \vec{r}_\mathrm{A}$, $\delta \vec{r}_\mathrm{B}$, $\delta \vec{r}_\mathrm{C}$, and $\delta \vec{r}_\mathrm{D}$ represent the directions in which the saddles are shifted, as calculated with eq.(\ref{e.ptb}).
}
\label{f.1}
\end{figure*}
		 
In order to grasp the general features of this model, we study several typical examples of the potential $U(x, y)$.
We first consider a simple potential function: $U(x, y) = - U_0 \cos(\pi x / \ell) \cos(\pi y / L)$ (see fig.~\ref{f.1}a), which we term Model I.
This potential possesses the reflection symmetry with respect to $x$: $U(x, y) = U(x_0 - x, y)$ with $^\exists x_0 \in [0, \ell)$.
Let $\vec{r}_\mathrm{m} = (\xmi, \ymi)$ denote a minimum of the energy function $V(x, y)$.
If the temperature is sufficiently low relative to the height of the barriers, $\Delta V$, the particle will hop from this minimum to one of its neighboring minima, $\vec{r}_\mathrm{m} + (\ell, L)$, $\vec{r}_\mathrm{m} + (\ell, - L)$, $\vec{r}_\mathrm{m} + (- \ell, L)$ or $\vec{r}_\mathrm{m} + (- \ell, - L)$, through the saddle $\vec{r}_\mathrm{A} = (\xmaa, \ymaa)$, $\vec{r}_\mathrm{B} = (\xmab, \ymab)$, $\vec{r}_\mathrm{C} = (\xmac, \ymac)$ or $\vec{r}_\mathrm{D} = (\xmad, \ymad)$, respectively (see fig.~\ref{f.1}a).
Because the two minima $\vec{r}_\mathrm{m} + (\ell, \pm L)$ located in the forward direction possess the same value of $x$, hops to them cannot be distinguished when we consider the particle motion projected onto the $x$ axis. The same is true for hops in the backward direction.
Note that the hopping motion of the particle becomes well described as a Markovian process in the low-temperature limit, even if one degree of freedom is not monitored. 
This is because whether the particle hops in the forward or backward direction becomes independent of all previous steps in the low-temperature limit.

Using multi-dimensional Kramers rate theory \cite{Kramers:1940, Landauer:1961, Langer:1969}, we find that the hopping rates from the well around $\vec{r}_\mathrm{m}$ to the wells around $\vec{r}_\mathrm{m} + (\ell, L)$, $\vec{r}_\mathrm{m} + (\ell, - L)$, $\vec{r}_\mathrm{m} + (- \ell, L)$ and $\vec{r}_\mathrm{m} + (- \ell, - L)$ are given by
\begin{eqnarray}
k_{(+, +)} &=& \frac{\lambda_\mathrm{A}}{2\pi}  \sqrt{| \det \mathcal{H}_\mathrm{m} \mathcal{H}_\mathrm{A}^{-1}|} ~ \e^{-\beta \left[ V(\vec{r}_\mathrm{A}) - V(\vec{r}_\mathrm{m}) \right]}, \label{e.k2d2++} \\
k_{(+, -)} &=& \frac{\lambda_\mathrm{B}}{2\pi} \sqrt{| \det \mathcal{H}_\mathrm{m} \mathcal{H}_\mathrm{B}^{-1}|} ~ \e^{-\beta \left[ V(\vec{r}_\mathrm{B}) - V(\vec{r}_\mathrm{m}) \right]} \label{e.k2d2+-}, \\
k_{(-, +)} &=& k_{(+, -)} \e^{-\beta (f \ell - F L)}, \label{e.k2d2-+} \\
k_{(-, -)} &=& k_{(+, +)} \e^{-\beta (f \ell + F L)}. \label{e.k2d2--}
\end{eqnarray}
Here, $\mathcal{H}_\mathrm{m}$, $\mathcal{H}_\mathrm{A}$ and $\mathcal{H}_\mathrm{B}$ denote the Hessian of $V(x, y)$ evaluated at $\vec{r}_\mathrm{m}$, $\vec{r}_\mathrm{A}$ and $\vec{r}_\mathrm{B}$, respectively. The values $\lambda_\mathrm{A}$ and $\lambda_\mathrm{B}$ are the positive eigenvalues of $-\sigma\mathcal{H}_\mathrm{A}$ and $- \sigma \mathcal{H}_\mathrm{B}$, where $\sigma$ is a matrix whose components are $\sigma_{11} = 1/\gamma$, $\sigma_{22} = 1/\Gamma$, and $\sigma_{12} = \sigma_{21} = 0$.
The periodicity of the system implies the equivalence of the saddles $\vec{r}_\mathrm{A}$ and $\vec{r}_\mathrm{D}$ and of the saddles $\vec{r}_\mathrm{B}$ and $\vec{r}_\mathrm{C}$ and that we have the relations $\xmad = \xmaa - \ell$, $\ymad = \ymaa - L$, $\xmac = \xmab - \ell$, and $\ymac = \ymab + L$. Then, eqs. (\ref{e.k2d2-+}) and (\ref{e.k2d2--}) follow.

The kinetic rates for forward and backward steps are given by $k_\pm = k_{(\pm, +)} + k_{(\pm, -)}$.
Therefore, from the definition of $\theta$ given in eq.~(\ref{e.mdb}), it can be expressed as
\begin{eqnarray}
\lefteqn{\theta = \frac{1}{\beta \ell} \pder{}{f} \ln \frac{k_+}{k_-}} \nonumber \\
&= \displaystyle \frac{\xmaa k_{(+, +)} + \xmab  k_{(+, -)}}{\ell [k_{(+, +)} + k_{(+, -)}]} - \frac{\xmac k_{(-, +)} + \xmad k_{(-, -)}}{\ell[ k_{(-, +)} + k_{(-, -)}]}. \label{e.theta}
\end{eqnarray}
Thus, it is found that the value of $\theta$ is given by an average of the $x$ coordinates of the saddles in which the weights are the hopping rates. This is one of the main results of this letter.

Because we are interested in the low-temperature limit, the hopping rates in the forward direction, $k_{(+, +)}$ and $k_{(+, -)}$, differ greatly in general. Let $\tilde x_+$ denote the $x$ coordinate of one of the two saddles in the forward direction at which the energy is lower. We term such a saddle a ``dominant saddle''.
The situation is similar for the hopping rates in the backward direction, $k_{(-, +)}$ and $k_{(-, -)}$.
Let $\tilde x_-$ denote the $x$ coordinate of the dominant saddle in the backward direction.
Then, in the limit $\beta \Delta V \to \infty$, we have
\begin{equation}
\theta = (\tilde x_+ - \tilde x_-)/\ell + O(T/\Delta V).
\label{e.simpletheta}
\end{equation}
 From this expression, it is found that the value of $\theta$ does not depend on the temperature if it is sufficiently low. This observation is consistent with recent experimental data \cite{Taniguchi:2005}.
It should be noted that in the case the dominant saddles are not uniquely determined (e.g., in the case $f = F = 0$ for Model I), eq.~(\ref{e.simpletheta}) is not applicable, while eq.~(\ref{e.theta}) remains valid.

Using eqs.~(\ref{e.theta}) and (\ref{e.simpletheta}), we can calculate the value of $\theta$ once the potential $U(x, y)$ is given. Figure \ref{f.1}b displays the result of our numerical calculation for Model I with $f = 0$.
It is observed that the value of $\theta$ is equal to unity when the driving force, $F$, is vanishing.
Moreover, it is found that $\theta$ is always unity if $F = 0$, not only for the case $f=0$ but also $f\neq 0$ (data not shown; see the discussion below).
However, if $F$ is switched on, the value of $\theta$ deviates from unity. In Model I, we find that $\theta$ decreases monotonically with the magnitude of $F$. We note that $\theta$ is an even function of $F$ because of the reflection symmetry of the system.
In fig.~1b, $\theta$ as a function of $F$, calculated through direct simulation of eqs.~(\ref{e.2d1}) and (\ref{e.2d2}) is plotted for various values of the temperature. It is seen that the form of $\theta$ seems to display the limiting behavior expressed in eq.~(\ref{e.simpletheta}) in the $T \to 0$ limit.

We now present arguments that account for the observations discussed above.
First, the reason that we have $\theta = 1$ when $F = 0$ can be understood as follows.
The transition probability of a microscopic path satisfies the LDB condition given in eq.~(\ref{e.ldb}), where $G(\vec{r}) \equiv V(x, y)$.
From this condition, by integrating out paths connecting $\vec{r}_1 = (x_0, y_0)$ and $\vec{r}_2 = (x, y)$, one can prove that the transition probability $P(x, y, t | x_0, y_0, t_0)$ satisfies the relation
\begin{equation}
\frac{P(x, y, t | x_0, y_0, t_0)}{P(x_0, y_0, t | x, y, t_0)} = \e^{\beta \left[ V(x_0, y_0) - V(x, y) \right]}.
\label{e.LDB2d}
\end{equation}

Because we observe the motion of the particle projected onto the $x$ axis, the relevant transition probability is the projected probability
\begin{equation}
\hat P(x, t | x_0, 0) \equiv \frac{\int \d y_0 \int \d y P_\mathrm{init} (x_0, y_0) P(x, y, t | x_0, y_0, 0)}{\hat P_\mathrm{init} (x_0)} ,
\label{e.ptp}
\end{equation}
where $P_\mathrm{init} (x, y)$ is the initial distribution and $\hat P_\mathrm{init} (x) \equiv \int \d y P_\mathrm{init} (x, y)$ represents a projected initial distribution.
Because the probability distribution relaxes inside a potential well very quickly before the occurrence of a hop, the projected transition probability is insensitive to the choice of the initial distribution if $t$ is much larger than the intra-well relaxation time, $\tau_\mathrm{in}$. Considering this point, we choose $P_\mathrm{init} (x, y) = C \exp[ - \beta U(x, y)]$, where $C$ is a normalization constant.

Then, using eqs.~(\ref{e.LDB2d}) and (\ref{e.ptp}), for $F = 0$, we obtain the expression
\begin{equation}
\hat P(\xmi + \ell, t | \xmi, 0) / \hat P(\xmi, t | \xmi + \ell, 0) = \e^{\beta f\ell},
\label{e.prm1}
\end{equation}
where the identity $\hat P_\mathrm{init} (\xmi) = \hat P_\mathrm{init} (\xmi + \ell)$ has been used.
Next, these transition probabilities are related to the kinetic rates as $\hat P(\xmi \pm \ell, \Delta t | \xmi, 0) = k_\pm \Delta t / \Omega + O(\Delta t^2)$ for $\Delta t$ satisfying $\tau_\mathrm{in} \ll \Delta t \ll (k_\pm)^{-1}$, where $\Omega \equiv \int \!\! \int_{\mathcal{A}} \d x \d y \exp[-\beta V(x, y)]$ and $\mathcal{A}$ denotes the interior of the potential well around $\vec r_{\rm m}$ \cite{Caroli:1981}.
Then, eq.~(\ref{e.prm1}) results in $k_+ / k_ - = \exp(\beta f \ell)$, and $\theta = 1$ is obtained for any value of $f$.

Next, we explain why $\theta$ is a decreasing function of $|F|$. The positions of the saddles, from which the value of $\theta$ is estimated according to eq.~(\ref{e.simpletheta}), are solutions of the fixed point equations
$\partial V(x, y)/ \partial x = 0$ and $\partial V(x, y)/ \partial y = 0$.
Let $\vec{r}_\mathrm{X} (\vec{E}) \equiv (x_\mathrm{X} (\vec{E}), y_\mathrm{X} (\vec{E}))$, where $\vec{E} \equiv (f, F)$, denote such a solution. When the force $\vec{E}$ is slightly modified as $\vec{E} \to \vec{E} + \vec{\delta E}$, where $\vec{\delta E} = (\delta f, \delta F)$, the fixed point moves as
\begin{equation}
\vec{r}_\mathrm{X} (\vec{E}) \to \vec{r}_\mathrm{X} (\vec{E} + \vec{\delta E}) = \vec{r}_\mathrm{X} (\vec{E}) + \mathcal{H}_\mathrm{X}^{-1} \vec{\delta E} + O(\vec{\delta E}^2),
\label{e.ptb}
\end{equation}
where $\mathcal{H}_\mathrm{X}$ is the Hessian of $V(x, y)$ evaluated at $\vec{r}_\mathrm{X} (\vec{E})$.

The shifts undergone by the saddle points due to a change in $\vec{E}$, as given by eq.~(\ref{e.ptb}), are depicted in fig.~\ref{f.1}c.
In the case of a positive driving force $(F \ge 0)$, the dominant saddles are $\vec{r}_\mathrm{A}$ and $\vec{r}_\mathrm{C}$.
In Model I, the eigenvector of $\mathcal{H}_\mathrm{A}$ at $\vec{r}_\mathrm{A}$ with a negative eigenvalue always lies in the first or third quadrant.
For this reason, eq.~(\ref{e.ptb}) leads to the inequality $\xmaa (\vec{E} + \vec{\delta E}) - \xmaa(\vec{E}) \le O(\vec{\delta E}^2)$ for $\delta f = 0$ and $\delta F \ge 0$. 
Analogously, we find $\xmac (\vec{E} + \vec{\delta E}) - \xmac(\vec{E}) \ge O(\vec{\delta E}^2)$. A slight increase of $F$ thus results in a decrease of $\theta$, in other words, we have $\partial \theta (\vec{\delta E})/ \partial F \le 0$ for $F \ge 0$.
Therefore, for $F \geq 0$ and fixed $f$, $\theta$ is a monotonically decreasing function of $F$.
A similar argument can be applied in the case of negative $F$, and we find that for $F \leq 0$ and fixed $f$, $\theta$ is a monotonically decreasing function of $|F|$.
In this way, the shift of the dominant saddles and, thus, the change in the value of $\theta$ are determined by the configurations of the eigenvectors at the saddles.

Summarizing the above arguments, in the case of Model I, we can conclude that $\theta$ decreases from unity as the magnitude of $F$ increases from zero.

		
\begin{figure*}
\onefigure{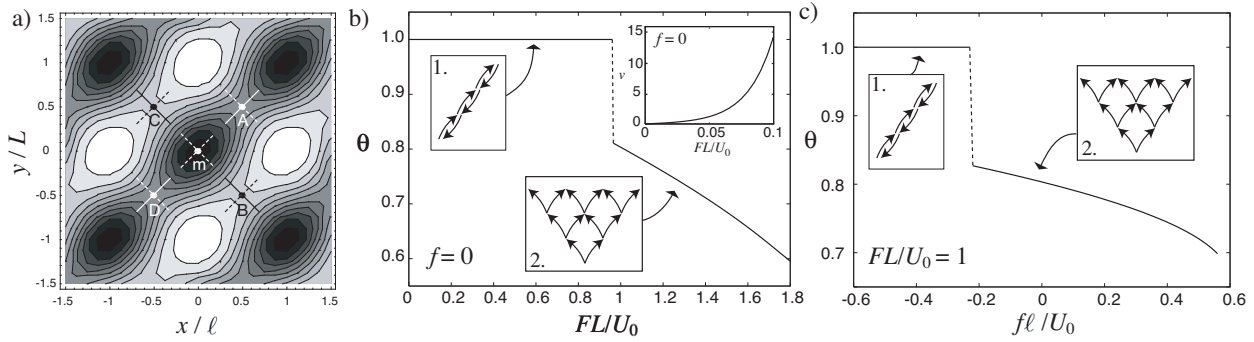}
\caption{a) Form of the potential $U(x, y) = - U_0  \left[ 1 + a \cos \pi \left(x/\ell - y/L \right) \right] \left[ 1 + \cos \left(\pi x /\ell \right) \cos \left(\pi y / L \right) \right]$, with $a = 0.5$ (Model II), represented by a grayscale, with the notation the same as in fig.~\ref{f.1}.
b) The reversibility parameter $\theta$ as a function of $F$ with $f = 0$. The solid curve represents the value of $\theta$ calculated using eq.~(\ref{e.simpletheta}). The structures of the dominant paths in the state space are depicted in insets 1 and 2 for each branch of the curve.
The upper inset displays the mean velocity as a function of the driving force $F$ with $f = 0$ for $\beta U_0 = 100$. The vertical axis is in units of $10^{-64} \kb T/(\gamma \ell)$.
c) $\theta$ as a function of $f$ with $F = U_0/L$ calculated using eq.~(\ref{e.simpletheta}).}
\label{f.2}
\end{figure*}

In order to check the generality of the above argument, we have investigated several other models.
Here, we consider one without the reflection symmetry with respect to $x$: $U(x, y) \neq U(x_0 - x, y)$ for $^\forall x_0 \in [0, \ell)$. The potential profile is plotted in fig.~\ref{f.2}a (to which we refer as Model II).
Because of the lack of the reflection symmetry in this model, the particle will realize a non-zero mean velocity in the $x$ direction even in the absence of the external force $f$, provided that the driving force $F$ is non-zero. This situation is reminiscent of that for a biological molecular motor (see the upper inset of fig.~\ref{f.2}b).

In fig~\ref{f.2}b, $\theta$ calculated using eq.~(\ref{e.simpletheta}) is plotted as a function of $F$ with $f=0$.
It is found that $\theta = 1$ over a certain range of values of $F$. Investigation of the potential at the saddles reveals that within this range of $F$, the dominant saddles are $\vec{r}_\mathrm{A}$ and $\vec{r}_\mathrm{D}$.
Therefore, by considering eq.~(\ref{e.simpletheta}) and the identity $x_\mathrm{D} = x_\mathrm{A} - \ell$, we conclude that $\theta = 1$.
Note that in this case, the dominant path for a forward step is the reverse of that for a backward step. Consequently, the trace of many sequential steps remains in a one-dimensional sub-space (see inset 1 of fig.~\ref{f.2}b).
Let us term this situation ``dominant-path reversible''.

However, at a certain critical value of $F$, denoted by $F_\mathrm{c}$, $\theta$ drops discontinuously from unity, and for $F > F_\mathrm{c}$, $\theta$ is a continuously decreasing function of $F$.
At $F = F_\mathrm{c}$, the roles of saddles are exchanged; the dominant saddle in the backward direction changes from $\vec{r}_\mathrm{D}$ to $\vec{r}_\mathrm{C}$.
Thus, for $F > F_\mathrm{c}$, the dominant saddles are $\vec{r}_\mathrm{A}$ and $\vec{r}_\mathrm{C}$, which become closer as $F$ increases for the same reason as that for Model I.
Then, from eq.~(\ref{e.simpletheta}), it is concluded that the value of $\theta$ decreases as $F$ increases.
In this case, the dominant path for a backward step is not the reverse of that for a forward step, and the trace of sequential steps forms a branched structure in the state space (see inset 2 of fig.~\ref{f.2}b). 
In contrast to the dominant-path-reversible situation, we term this situation ``dominant-path irreversible''.
Note that a similar phenomenon occurs for negative $F$ because of the symmetry of the system.
Moreover, this type of switching also occurs when the load, $f$, is varied with fixed $F$, as plotted in fig.~\ref{f.2}d.

In this way, the value of $\theta$ directly reflects the structure of the dominant paths in the state space.
It is interesting that the conventionally analyzed quantities such as the mean velocity and the diffusion coefficient do not exhibit such a sharp transition.
Thus, it is the value of $\theta$, not these other quantities, that provides the information of the structure of the dominant paths in the state space of the system.


In conclusion, the physical significance of the dimensionless parameter $\theta$, which has been defined as the ratio of the characteristic distance to the actual step size of a molecular motor, has been investigated by considering simple Langevin models that have a hidden degree of freedom. We have identified the mechanism through which $\theta$ deviates from unity in these models.
It was found that if no driving force is applied to the hidden degree of freedom, $\theta$ is always unity.
This means that if we find $\theta \neq 1$ in a certain system, there could be at least one hidden degree of freedom that is directly coupled to an external energy source.
Furthermore, $\theta$ is always unity even when a driving force is present,
if the system is in ``dominant-path reversible'' situation, i.e., the dominant path for a forward step is the time-reverse of the dominant path for a backward step. 
By contrast, $\theta$ deviates from unity if the system is in ``dominant-path irreversible'' situation, i.e., the dominant path for a forward step is not the time-reverse of the dominant path for a backward step.
In this case, the dominant paths have a branched structure in the state space.
Thus, the value of $\theta$ serves as an indicator of the structure of the dominant paths in the state space even when several degrees of freedom are not monitored.
These features of $\theta$ are expected to be valid if the number of hidden degree of freedom exceeds one, since the main results presented in this letter are easily extended to systems with more degree of freedom.

On the basis of the above results, we now present an interpretation of relevant experimental findings for molecular motors with regard to dominant-path reversibility.
As noted above, it has been found that most of biological motors exhibit values of $\theta$ less than unity.
According to our discussion, there could be hidden degrees of freedom that are coupled to the energy source (i.e. the reservoir of ATP, ADP and Pi) and that the dominant paths in the state space are branched.
This seems to imply that these motors in general hydrolyze ATP molecules even in backward steps (dominant-path irreversible).
This is consistent with several experimental results on conventional kinesin \cite{Nishiyama:2002, Carter:2005}.

In Model II, there was a situation where $\theta$ remains unity while non-zero velocity is exhibited (dominant-path reversible).
In this case, the dominant path for a backward step is the reverse of the dominant path for a forward step, which implies that ATP is synthesized when the motor steps backward.
This situation could correspond to the case of F$_1$ ATP synthase that synthesizes ATP when it is forced to rotate backward \cite{Itoh:2004, Rondelez:2005}.
Therefore, it is expected from our argument that the value of $\theta$ for this molecule is close to unity.
Although the value of $\theta$ has not been determined for this molecule, its determination will serve as an experimental test of the present theory.

To properly interpret the results of this work, it should be noted that the connection between the models studied here and actual molecular motors is not clear, as these models are quite simple.
For instance, the actual molecular motors are expected to have several substates those are not necessarily coupled to its stepwise motion. Since such substates can also be incorporated to the model studied in this letter by modifying the profile of the potential $U(x, y)$, one would be able to see the effects of such substates on the property of the characteristic distance.
Moreover, in most models we have studied, including two models exemplified here, the value of $\theta$ is less than or equal to unity. It is not clear at present whether there is a case in which $\theta$ is greater than unity.
Further theoretical and experimental studies of the parameter $\theta$ should lead to a better understanding of the mechanisms and architectures of biological molecular motors.

\acknowledgments
The authors acknowledge helpful discussions with T. S. Komatsu and S. -i. Sasa.
This work was partly supported by MEXT KAKENHI (No. 16740217).



\begin{thebibliography}{99}

\bibitem{Nishiyama:2002} Nishiyama M., Higuchi H. and Yanagida T., Nature Cell Biol., \textbf{4} (2002) 790.

\bibitem{Carter:2005} Carter N. J., Cross R. A., Nature \textbf{435} (2005) 308.

\bibitem{Taniguchi:2005} Taniguchi Y., Nishiyama M., Ishii Y. and Yanagida T., Nature Chem. Biol., \textbf{1} (2005) 342.

\bibitem{Hirakawa:2000} Hirakawa E., Higuchi H. and Toyoshima Y. Y., Proc. Natl. Acad. Sci. USA, \textbf{97} (2000) 2533.

\bibitem{Clemen:2005} Clemen A. E.-M., Vilfan M., Jaud J., Zhang J., B\"armann M. and Rief M., Biophys. J., \textbf{88} (2005) 4402.

\bibitem{Hill:1977} Hill T. L., \textit{Free Energy Transduction in Biology} (Academic Press) 1977.

\bibitem{Maes:1999} Maes C., J. Stat. Phys., \textbf{95} (1999) 367.

\bibitem{Crooks:2000} Crooks G. E., Phys. Rev. E, \textbf{61} (2000) 2361.

\bibitem{Harada:2006} Harada T. and Sasa S.-i., Phys. Rev. E, \textbf{73} (2006) 026131.

\bibitem{Jarzynski:2000} Jarzynski C., J. Stat. Phys., \textbf{98} (2000) 77.

\bibitem{Fisher:1999} Fisher M. E., Kolomeisky A. B., Proc. Natl. Acad. Sci. USA, \textbf{96} (1999) 6597.

\bibitem{Kolomeisky:2005} Kolomeisky A. B., Stukalin E. B. and Popov A. A., Phys. Rev. E, \textbf{71} 031902 (2005).

\bibitem{Seifert:2005} Seifert U., Europhys. Lett., \textbf{70} (2005) 36.

\bibitem{Julicher:1997} J\"ulicher F., Ajdari A. and Prost J., Rev. Mod. Phys., \textbf{69} (1997) 1269.

\bibitem{Reimann:2002} Reimann P., Phys. Rep., \textbf{361} (2002) 57.

\bibitem{Bell:1978} Bell G. I., Science, \textbf{200} (1978) 618.

\bibitem{Magnasco:1994} Magnasco M. O., Phys. Rev. Lett., \textbf{72} (1994) 2656.

\bibitem{Terada:2002} Terada T. P., Sasai M. and Yomo T., Proc. Natl. Acad. Sci. USA, \textbf{99} (2002) 9202.

\bibitem{Onsager:1953} Onsager L. and Machlup S., Phys. Rev., \textbf{91} (1953) 1505.

\bibitem{Kramers:1940} Kramers H. A., Physica, \textbf{7} (1940) 284.

\bibitem{Landauer:1961} Landauer R. and Swanson J. A., Phys. Rev., \textbf{121} (1961) 1668.

\bibitem{Langer:1969} Langer J. S., Ann. Phys. (N.Y.) \textbf{58} (1969) 258.

\bibitem{Caroli:1981} Caroli B., Caroli C. and Roulet B., J. Stat. Phys. \textbf{26} (1981) 83.

\bibitem{Itoh:2004} Itoh H., Takahashi A., Adachi K., Noji H., Yasuda R., Yoshida M. and Kinosita K., Nature, \textbf{427} (2004) 465.

\bibitem{Rondelez:2005} Rondelez Y., Tresset G., Nakashima T., Kato-Yamada Y., Fujita H., Takeuchi S. and Noji H., Nature, \textbf{427} (2005) 773.

\end{thebibliography}
\end{document}